\documentclass[journal,comsoc]{IEEEtran}

\usepackage[T1]{fontenc}% optional T1 font encoding

\usepackage{xcolor}
\usepackage{cite}
\usepackage{graphicx}

\usepackage{amsmath}
\DeclareMathOperator{\diag}{diag}
\usepackage[cmintegrals]{newtxmath}

\usepackage[caption=false,font=footnotesize]{subfig}

\usepackage{xcolor}

\begin{document}

\title{\huge Reconfigurable Intelligent Surfaces for Wireless Communications: Principles, Challenges, and Opportunities}

\author{
        Mohamed A. ElMossallamy,~\IEEEmembership{Student Member,~IEEE,}
        Hongliang Zhang,~\IEEEmembership{Member,~IEEE,}
        Lingyang Song,~\IEEEmembership{Fellow,~IEEE,}
        Karim G. Seddik,~\IEEEmembership{Senior Member,~IEEE,}
        Zhu Han,~\IEEEmembership{Fellow,~IEEE,} \\ and 
        Geoffrey Ye Li,~\IEEEmembership{Fellow,~IEEE}
        
\thanks{Mohamed A. ElMossallamy, Hongliang Zhang, and Zhu Han are with the Electrical and Computer Engineering Department, University of Houston, TX, USA. Zhu Han is also with the Department of Computer Science and Engineering, Kyung Hee University, Seoul, South Korea (emails: m.ali@ieee.org, hongliang.zhang92@gmail.com, zhan2@uh.edu).}

\thanks{Lingyang Song is with Department of Electronics, Peking University, Beijing, China (email: lingyang.song@pku.edu.cn).}% <-this % stops a space

\thanks{Karim G. Seddik is with the Electronics and Communications Engineering Department, American University in Cairo, New Cairo, Egypt (email: kseddik@aucegypt.edu).}

\thanks{Geoffrey Ye Li is with the School of Electrical and Computer Engineering, Georgia Institute of Technology, Atlanta, GA, USA (email: liye@ece.gatech.edu).}

}

\maketitle

\begin{abstract}
Recently there has been a flurry of research on the use of reconfigurable intelligent surfaces (RIS) in wireless networks to create smart radio environments. In a smart radio environment, surfaces are capable of manipulating the propagation of incident electromagnetic waves in a programmable manner to actively alter the channel realization, which turns the wireless channel into a controllable system block that can be optimized to improve overall system performance. In this article, we provide a tutorial overview of reconfigurable intelligent surfaces (RIS) for wireless communications. We describe the working principles of reconfigurable intelligent surfaces (RIS) and elaborate on different candidate implementations using metasurfaces and reflectarrays. We discuss the channel models suitable for both implementations and examine the feasibility of obtaining accurate channel estimates.  Furthermore, we discuss the aspects that differentiate RIS optimization from precoding for traditional MIMO arrays highlighting both the arising challenges and the potential opportunities associated with this emerging technology. Finally, we present numerical results to illustrate the power of an RIS in shaping the key properties of a MIMO channel.
\end{abstract}

\section{Introduction}
\IEEEPARstart{I}{n} the relentless pursuit to increase the capacity of wireless networks and support higher and higher data rates, wireless system designers have sought to optimize every aspect of communication systems. This has started with more spectrum efficient waveforms and multiplexing techniques (e.g. OFDM), leveraging the spatial domain (e.g. MIMO), and inching ever closer to the maximum theoretical capacity using more advanced adaptive modulation and coding techniques. On the network side, cellular networks have gotten denser with more aggressive frequency reuse and inter-cell coordination techniques have been developed to deal with the resulting interference. However, the capacity of the network is still ultimately limited by the unreliability of wireless propagation and the available spectrum. 

To tackle the shortage of spectrum, communications systems have been steadily moving into higher frequency bands where amble unused spectra exist. However, the unreliable stochastic nature of wireless propagation remains inevitable. Conventional wisdom regards the wireless channel as an uncontrollable stochastic link with inherent unreliability. Therefore, the best we can do is to understand it, model it, and combat its unpredictability with sophisticated signal processing at the transmitter and the receiver. This traditionally includes diversity techniques, beamforming, and adaptive coding and modulation to squeeze as much usable capacity as possible. Recently, with the advent of reconfigurable intelligent surfaces (RIS) and the emergence of the concept of the smart radio environment\cite{renzo_smart_2019, liang_lisa_2019}, we might be able to also control, at least partially, the wireless channel itself.

The idea of controlling the ambient environment to provide more favorable propagation characteristics represents a paradigm shift in how we think about wireless systems design. Instead of treating reflection and scattering in the environment as uncontrollable phenomena whose effects can only be modeled stochastically, they become part of the system parameters that may be optimized, which can overcome many of the challenges of wireless communications.

\begin{figure}[tbp]
  \centering
  \includegraphics[height=6.5cm]{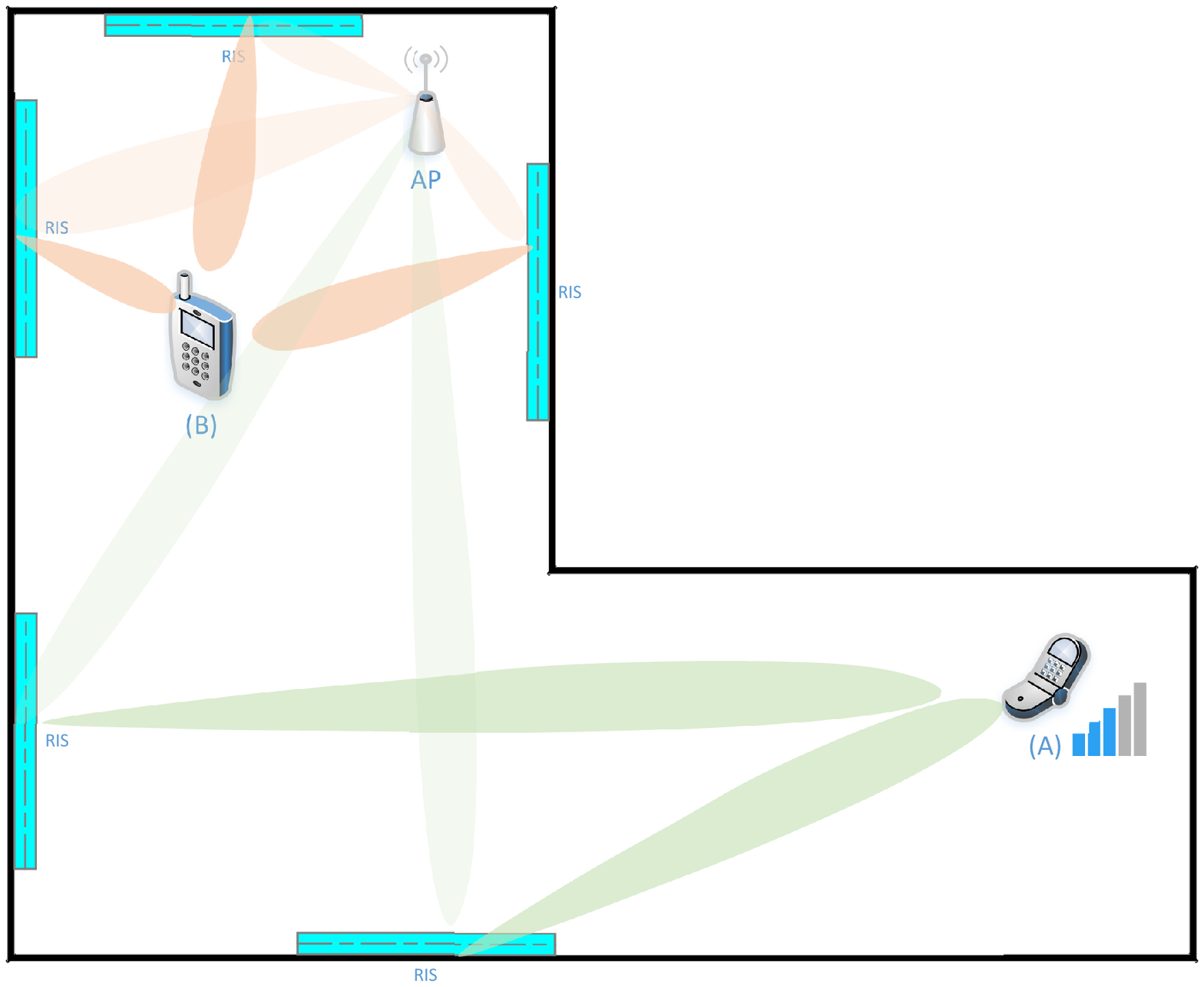} 
  \caption{A smart radio environment with multiple RISs. User A is far away from the AP and suffers from low received signal strength, while user B has amble received power but a low-rank ill-conditioned channel. The RISs can be optimized to help in both scenarios.}
  \label{fig:smartenv}
\end{figure}

In general, the practically achievable rate over a wireless link is limited by the order of the modulation and the number of spatial streams. Both are decided according to the current channel realization. The order of modulation is adapted according to the signal strength perceived at the receiver, which is a consequence of the \emph{channel gain}. To keep error rates low and avoid re-transmissions, a user at the cell edge will be forced to use lower order modulation, and thus suffer from lower rates. On the other hand, the number of spatial streams is adapted according to the number of the usable eigenmodes of the channel. A line-of-sight (LOS) link may enjoy high channel gain, but will probably suffer from a \emph{spatially-sparse} \emph{low-rank channel}, limiting the number of spatial streams and consequently the achievable rate. The two scenarios are illustrated in Fig.~\ref{fig:smartenv}. These scenarios can arise in any wireless network, but future communication systems are expected to be especially affected. In particular, the propagation characteristics at the higher frequency bands used in future generations of communications systems, e.g. 30-100 GHz, will give rise to these scenarios more often. Reconfigurable intelligent surfaces can be used to alter the channel realization in these scenarios and improve overall system performance dramatically.

Wireless communication systems have traditionally relied on frequency bands extending from a few hundred megahertz to a few gigahertz. This was mainly driven by the favorable propagation characteristics at these bands and the ease of implementing efficient and cheap transceivers. Although there exists a huge swath of spectrum at the millimeter-wave range (30-100 GHz), it has been traditionally sought that these frequencies are not suitable for wireless communication, especially outdoor cellular communications. Subsequent intensive measurement campaigns \cite{akdeniz_millimeter_2014} have shown the potential of millimeter-wave bands when combined with directional high gain antenna arrays. This has led to immense interest in millimeter wave wireless communications.

This foray into higher frequency bands necessitated a reevaluation of channel modeling techniques. Communicating at millimeter-wave frequencies poses a lot of challenges. The channel at these frequencies is significantly more hostile than at sub 6 GHz frequencies. Diffraction ceases to be a reliable propagation mechanism with line-of-sight, first-order reflections and scattering becoming much more dominant. This means shadowing will have severe detrimental effects on the average received power. Indeed, channel models developed for millimeter-wave include a third state, in addition to line-of-sight and non-line-of-sight, to explicitly model an outage event when received power is too weak to establish a link \cite{akdeniz_millimeter_2014}. Although adaptive beam steering techniques can improve the reliability of millimeter-wave links, communications at these frequencies remain very challenging.

\begin{figure*}[!t] 
  \subfloat[A 48-element reflectarray-based RIS. Each element is a traditional antenna connected to a phase shifter.]{% 
    \includegraphics[height=6cm]{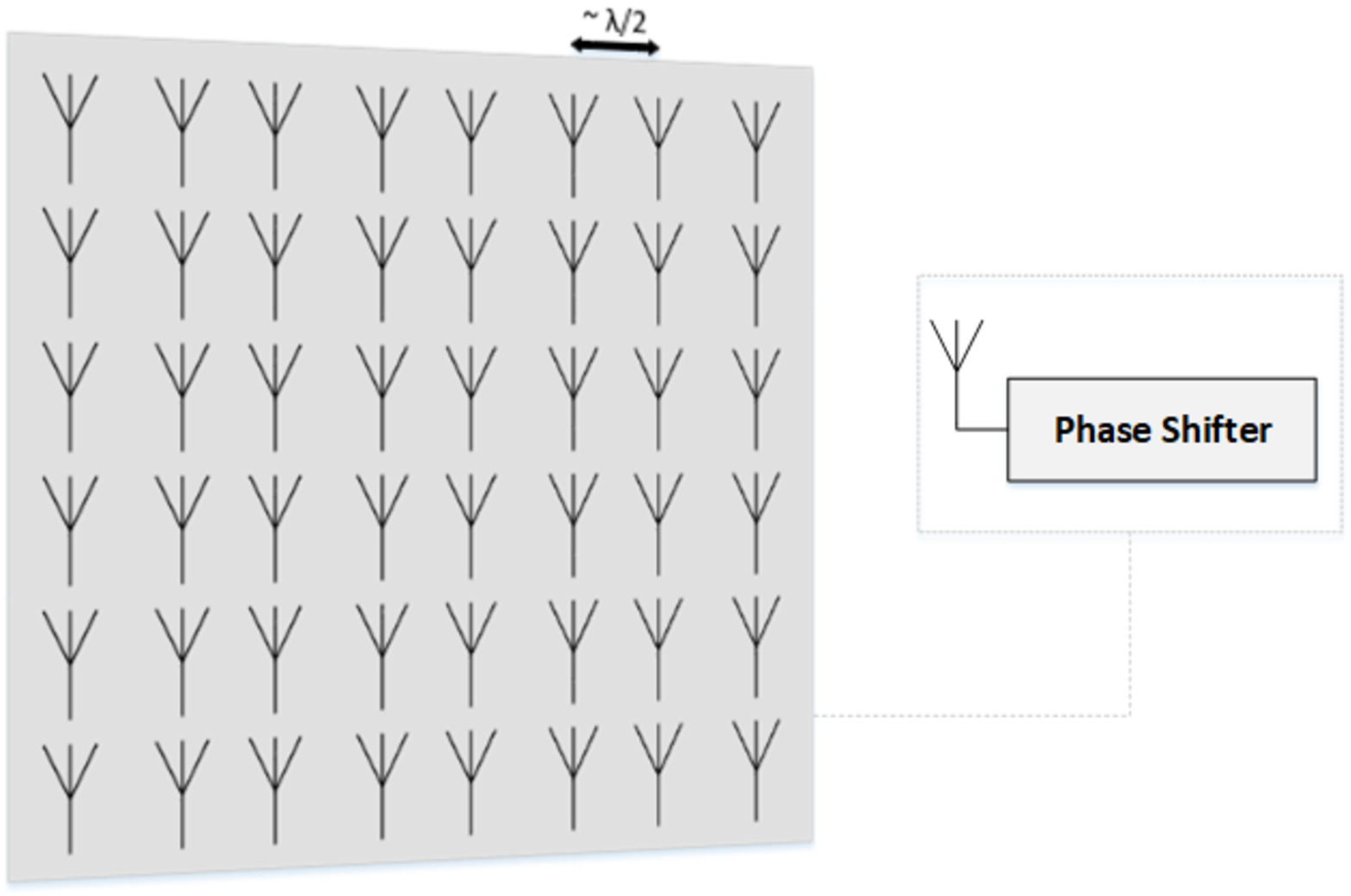} 
  } 
  \hfill 
  \subfloat[A 4-element metasurface-based RIS. Each element/tile is a dynamic metasurafce with numerous tightly-packed meta-atoms and can apply an arbitrary quasi-continuous phase gradient.]{% 
    \includegraphics[height=6cm]{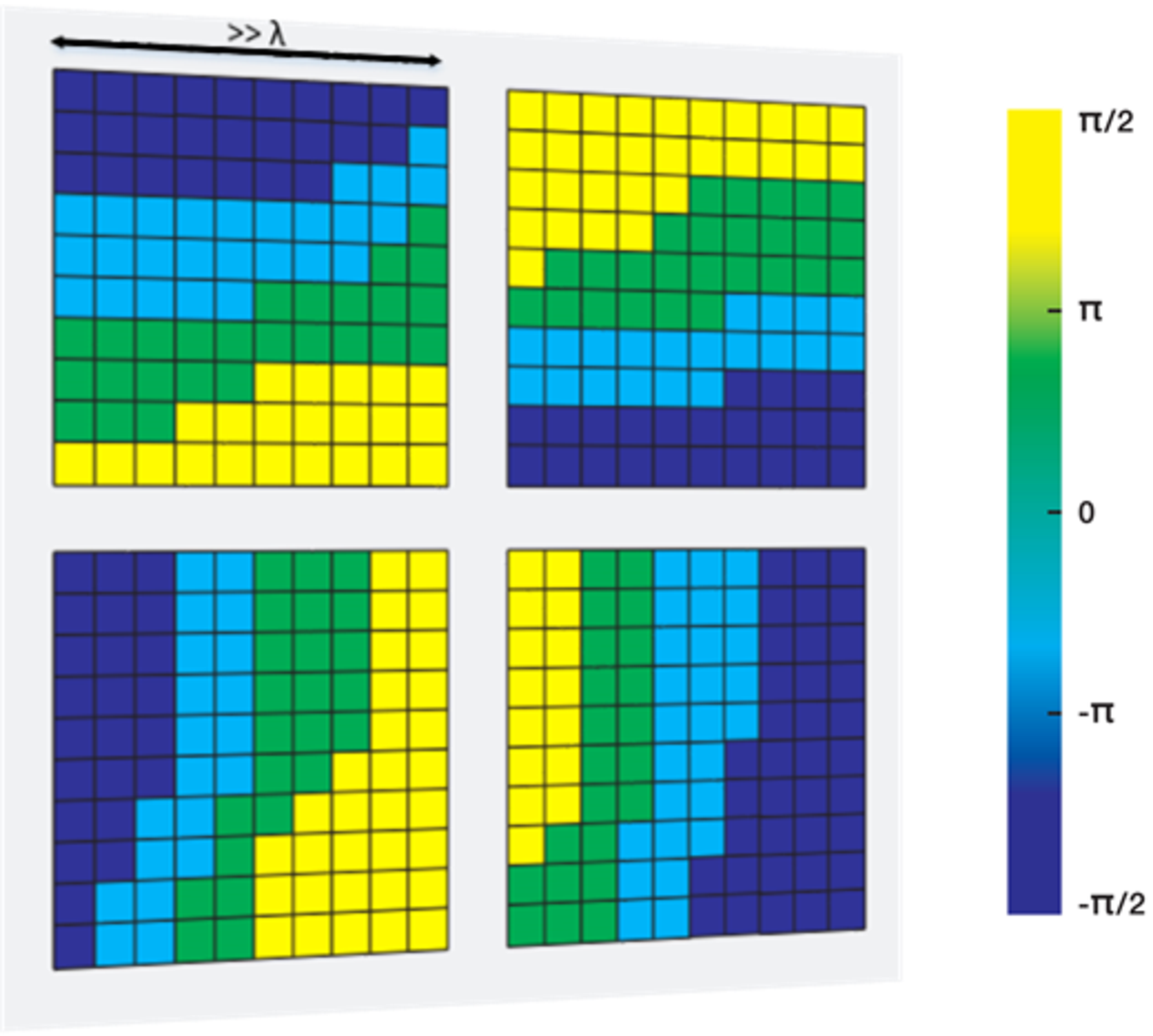} 
  } 
  \centering
  \caption{Different Implementations of an RIS.} 
  \label{fig:implementations}
\end{figure*}

Another challenge on millimeter-wave channels is that they are spatially sparse which means that only a small number of propagation paths exist between the transmitter and the receiver. This is in contrast to the rich scattering assumption usually utilized in sub 6 GHz channels. Although this sparsity could be leveraged in channel estimation and precoding, especially for hybrid analog/digital architectures \cite{ayach_spatially_2014}, it also limits the number of spatial data streams that can be supported by the channel. Hence, the spatial multiplexing capability of millimeter-wave channels is limited. In many cases, e.g. LOS, only a single viable propagation path exists and spatial multiplexing is not feasible. 

The above challenges can be overcome by leveraging the power of RISs. In low-received power scenarios, the RISs can function as centralized beamformer to increase channel gains or create a propagation path around major obstacles to restore a link in an outage. While in LOS spatially-sparse scenarios, the RIS can be used to emulate a rich scattering environment to enhance the channel condition number and improve spatial multiplexing capability. Different technologies can be used to implement an RIS \cite{hum_reconfigurable_2014, he_tunable_2019}. In its most simple form, an RIS can be implemented as a dynamic reflectarray \cite{hum_reconfigurable_2014}, whose elements are omnidirectional antennas with controllable termination that can be changed dynamically to backscatter and phase shift the incident waveform. A more elaborate implementation would be using a dynamically tunable metasurface\cite{he_tunable_2019}, a 2D planar form of metamaterials that have been shown to possess great electromagnetic wave manipulation capabilities. Relying on the metasurface implementation, an RIS element can not only scatter and phase-shift the signal but can also act as an anomalous mirror with a controllable reflection angle and even polarization manipulation abilities. The implementation can also greatly affect the link-budget as reflected and scattered waves suffer from different path loss scaling. However, as the RIS element gains more wave manipulation capabilities, the optimization task gets more complicated.

In this article, we provide a comprehensive tutorial on reconfigurable intelligent surfaces for wireless communications. We first discuss the different implementations of reconfigurable intelligent surfaces and elaborate on their differences with respect to wave manipulation capabilities, channel modeling, and the link budget. We then discuss the challenges of optimizing the RIS parameters to benefit the overall system performance, and also highlight the potential opportunities if the challenges can be overcome. Finally, we provide numerical results to illustrate the power of an RIS in shaping some key properties of a MIMO channel to enhance performance and simplify transceiver designs.

The rest of this article is organized as follows. In Section~\ref{Sec::RIS}, we introduce different implementations to realize reconfigurable intelligent surfaces. In Section~\ref{Sec::ChanModel}, we discuss how to incorporate the effects of the RISs in the channel model, how the implementation technology can influence the choice of the appropriate model, and how different implementations can affect the link budget. In Section~\ref{Sec::Optimization}, we review the state-of-the-art of RIS-assisted system optimization techniques and discuss the difference between the optimization of RIS configurations and precoding for traditional MIMO arrays highlighting both the arising challenges and the potential opportunities associated with this emerging technology. In Section~\ref{Sec::Results}, we present numerical results showing the power of RIS in improving the spatial multiplexing capability of MIMO channels and simplifying transceivers. Finally, we identify some future directions in Section~\ref{Sec::FutDirec} and conclude the article in Section~\ref{Sec::Conclusion}.

\section{Reconfigurable Intelligent Surfaces}\label{Sec::RIS}

In a smart radio environment \cite{renzo_smart_2019}, shown in Fig.~\ref{fig:smartenv}, one or more RISs can be used to influence wireless propagation in a manner that is beneficial to the overall system performance. For example, by increasing received power through beamforming or by influencing the channel rank and condition number to facilitate spatial multiplexing.  In essence, any passive surface that can be dynamically reconfigured to manipulate incident electromagnetic waves and change the channel conditions change can be called an RIS. This definition is valid regardless of the particular implementation. Two main implementations have been investigated in the literature; based on traditional reflectarrays, or metasurfaces. Regardless of the implementation, an RIS should be passive, that is it does not emit any power of its own and only aims to manipulate existing transmitted waves. In this aspect, RIS is similar to backscatter technology and different from relaying. In this section, we elaborate on these RIS implementations.

\subsection{Reflectarray-based Implementation}

As shown in Fig.~\ref{fig:implementations}, the simplest way to implement a reconfigurable intelligent surface is to use a passive reflectarray whose elements' antenna termination can be controlled electronically to backscatter and phase-shift the incident signal\cite{hum_reconfigurable_2014}. Each element individually has a very limited effect on the propagated waves, but a sufficiently large number of elements can effectively manipulate the incident wave in a controllable manner. To be effective, this implementation would require a vastly large number of antenna elements, probably thousands \cite{arun_rfocus:_2019}. 

Each element in the reflectarray-based RIS is similar to a tag in backscatter communications systems. However, there are two main differences. The first one is that the reflections are used to communicate information from the reflector to the receiver in backscatter communications, while the RIS aims only to help the ongoing transmission and does not communicate information of its own in an RIS-assisted communication scenario. The other one is the size and scale of the RIS. Although a single element in a reflectarray-based RIS is similar to a backscatter tag, the elements of an RIS work collectively, based on knowledge of the propagation environment, over a very large area to induce significantly more potent effects on the incident waves.

The reflectarray-based RIS can be thought of as providing powerful centralized analog beamforming capabilities in advantageous locations that can be utilized by communication endpoints. This can also lead to much simpler transmitters and receivers by shifting complexity to the RIS and controllers. Finally, note that the dimension of each element in a reflectarray-based RIS is comparable to the wavelength, e.g. $\frac{\lambda}{2}$, and individually act as a diffuse scatterer.

\subsection{Metasurface-based Implementation}

A more sophisticated implementation of an RIS might be done using metasurfaces\cite{chenmetareview16}. A metasurface is the two-dimensional planar form of metamaterials, which are man-made synthetic materials with electromagnetic properties not found in naturally occurring materials. They were originally conceived for applications in the optical domain to allow cheap and robust planar optical components replacing more expensive custom-made lenses.

A metasurface is comprised of a large number of closely spaced deeply subwavelength resonating structures called pixels or meta-atoms\cite{chenmetareview16}. \emph{Both individual meta-atoms and the space between adjacent meta-atoms are much smaller than the wavelength in size.} The very small size of these closely-packed atoms and their large number offer a vast number of degrees of freedom in manipulating the incident electromagnetic waves. In particular, by deliberately designing its meta-atoms, a metasurface can impose arbitrary quasi-continuous \cite{huang_holographic_2019} amplitude/phase profiles on the incident wave-fronts and exercise fine-grained control over the scattered electric field.

Earlier metasurfaces designs have been based on static preset meta-atoms designs that cannot be modified after fabrication, which is good enough for making custom lenses for optical applications. However, later designs rely on semiconductor components, which can be reconfigured in real-time to change the underlying meta-atom structure and hence the electromagnetic behavior of the metasurface \cite{he_tunable_2019}. This reconfigurability is achieved by integrating components that can be tuned either electrically, mechanically, or thermally. Electrically tunable metasurfaces are especially attractive since they can be cheaply manufactured using well-understood semiconductor technologies, and can be tuned fast enough to adapt to the time-varying wireless channel. For instance, by incorporating varactor diodes or liquid crystals within the meta-atoms. This dynamic tunability is paramount in wireless applications to allow adapting to the changing channel realization. 

A metasurface-based RIS is comprised of several tiles, where each tile is an individually reconfigurable metasurface whose dimensions are much larger than the wavelength. In a sense, each individual element in a metasurface-based RIS has functionality akin to a reflectarray on its own right. In particular, you can consider each tile in a reconfigurable metasurface as the limit of a reflectarray as both the antenna size and antenna spacing diminish and the applied amplitude/phase profile becomes approximately continuous along the surface. This allows a large degree of flexibility in manipulating the incident wave-front. For example, each tile can reflect the incident wave-front in a different direction. However, the majority of empirical works in the literature \cite{tan_increasing_2016, welkie_programmable_2017, arun_rfocus:_2019, hougne_optimally_2019, li_towards_LAIA} employ reflectarray-based RIS. Even with the simpler reflectarray-based implementation, impressive results have been reported. In theory, the increased flexibility associated with the metasurface-based RIS leads to even better results; however, only real-world experimentation will tell if the increased sophistication would lead to practical performance gains.

\section{Channel Model}\label{Sec::ChanModel}
\begin{figure*}[t] 
  \subfloat[Large-scale path loss of diffusely scattered electromagnetic waves.]{% 
    \includegraphics[height=6cm]{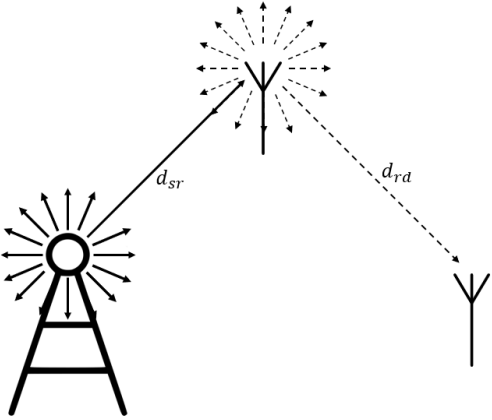} 
  } 
  \hfill 
  \subfloat[Large-scale path loss of reflected electromagnetic waves.]{% 
    \includegraphics[height=6cm]{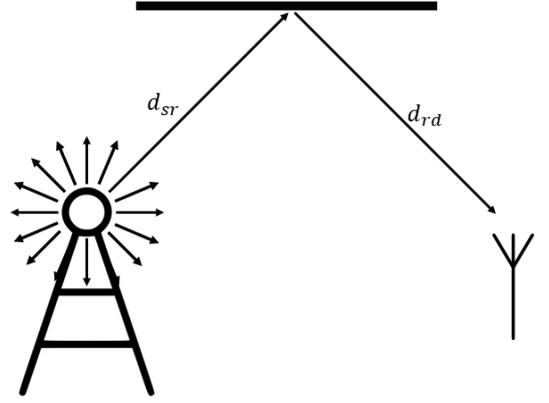} 
  } 
  \centering
  \caption{Implementation effects on large-scale pathloss.} 
  \label{fig:pathloss}
\end{figure*}

In this section, we discuss how to incorporate the existence of an RIS in the propagation environment into the channel model. Accurate channel models are essential for the analytical study of RIS-assisted communications as well as simulations. This is paramount in the evaluation of the utility of the RIS, especially as it compares to other technologies, such as relaying. The implementation method of the RIS, whether its reflectarrays or metasurfaces, may play a role in deciding which modeling technique is more appropriate. We present two modeling techniques used in the literature and discuss their underlying assumptions and rationale.

Consider a scenario where an $M$-antenna transmitter communicates with an $N$-antenna receiver and an $L$-element RIS exists in the environment. Assuming a narrowband flat-fading scenario, the received signal can be written as
\begin{equation}\label{eq:rcvdSig_Cascaded}
\begin{split}
\mathbf{y} &= \sqrt{G_{d}} ~ \mathbf{H}_{\mathrm{env}} \mathbf{x} + \sqrt{G_{r}} ~ \mathbf{H}_{\mathrm{RIS}} \mathbf{x} +  \mathbf{n},\\
           &= \mathbf{H}_{\mathrm{eff}} \mathbf{x} +  \mathbf{n},
\end{split}
\end{equation}
where $\mathbf{x}$ is the transmitted vector, $\mathbf{n}$ is the white Gaussian noise vector at the $N$ receiver's antennas. $\mathbf{H}_{\mathrm{env}}$ represents the uncontrollable channel between the transmitter and the receiver, excluding the effects of the RIS, and $\mathbf{H}_{\mathrm{RIS}}$ represents the controllable channel between the transmitter and the receiver through the RIS. $\sqrt{G_{d}}$ and $\sqrt{G_{r}}$ represent the large-scale gain associated with $\mathbf{H}_{\mathrm{env}}$ and $\mathbf{H}_{\mathrm{RIS}}$, respectively. Finally, $\mathbf{H}_{\mathrm{eff}} \coloneqq \sqrt{G_{d}}~ \mathbf{H}_{\mathrm{env}} + \sqrt{G_{r}}~ \mathbf{H}_{\mathrm{RIS}}$ represents the overall effective RIS-augmented channel seen by the transceivers.

It is worth discussing how the effective channel in RIS-assisted communications differs from the effective channel in traditional precoded multi-antenna systems. In traditional precoded systems, the equivalent channel is given by the product $\mathbf{H}_{\mathrm{eff}} = \mathbf{H}_{\mathrm{env}} \mathbf{P}$, where $\mathbf{P}$ is the applied precoder. While in RIS-assisted communications, the RIS effect is additive, cf.~\eqref{eq:rcvdSig_Cascaded}. Furthermore, elements of a traditional precoder can take any value satisfying some constraints, e.g. power, while the additive term in RIS effective channel depends on the propagation environment and is only partially controllable via the RIS. Although this makes the RIS optimization task more challenging in general, the additive effect of the RIS is more potent than the mixing and steering effects of the traditional multiplicative precoders \cite{heathLozano_foundations_2018}, giving the system designer more control over the effective channel. Next, we present two techniques to model the channel through the RIS, $\mathbf{H}_{\mathrm{RIS}}$.

\subsection{Dyadic Backscatter Channel Model}
The first technique models the RIS at the scatterer level and assumes each element in the RIS is a regular omnidirectional antenna subject to the effects of fading and uses the so-called dyadic backscatter channel \cite{griffin_gains_2008} to model the channel through the RIS. Using this model, the channel through the RIS, $\mathbf{H}_{\mathrm{RIS}}$, can be written as
\begin{equation}\label{eq:ris_dbc_channel}
\begin{split}
\mathbf{H}_{\mathrm{RIS}} = \mathbf{F} \mathbf{Q} \mathbf{G},
\end{split}
\end{equation}
where $\mathbf{F}$ is the $N \times L$ channel from the RIS to the receiver, $\mathbf{G}$ is the $L \times M$ channel from the transmitter to the RIS. The $L \times L$ matrix, $\mathbf{Q}$, represents the interaction of the RIS with the transmitted waveform. Assuming no coupling between the RIS elements, the interaction matrix can be written as
\begin{equation}\label{eq:interactionMat_DBC}
\begin{split}
\mathbf{Q} = \diag \left( \beta_1 \mathrm{e}^{i \theta_1}, \beta_2 \mathrm{e}^{i \theta_2}, \dots, \beta_L \mathrm{e}^{i \theta_L} \right),
\end{split}
\end{equation}
where $\beta_i \in \left[0,1\right]$ and $\theta_i \in \left[0, 2\pi\right)$ and can be controlled by changing the complex antenna load similar to backscatter tags \cite{elmossallamy_noncoherent_2019}. Note that the phase shifts can either be continuous \cite{tan_increasing_2016}, or discrete \cite{hzhang_reconfigurable_2019,  hum_reconfigurable_2014} based on implementation, e.g., varactors vs. switched loads.

From~\eqref{eq:ris_dbc_channel}, for any assumed statistical distribution for the entries of $\mathbf{G}$ and $\mathbf{F}$, the overall channel distribution will be given by their product distribution. In general, this kind of cascaded fading is known to have more detrimental effects on performance compared to regular fading. However, increasing the number of backscattering elements, $L$, does improve the fading characteristics \cite{griffin_gains_2008}. More importantly,  it is probably not an adequate model for the metasurface-based RIS, where each element is not a typical antenna but rather a planar surface with dimensions much larger the wavelength. Hence, typical statistical characterization of the received signal envelope by an antenna, i.e. fading, may not be appropriate.

\subsection{Spatial Scattering Channel Model}

Now, we move to the second channel modeling technique, which avoids some of the limitations of the cascaded dyadic backscatter model presented above, and better reflects the propagation mechanisms through a metasurface-based RIS. Assuming each element in the RIS is much larger than the wavelength, we can model each element as a reflector in the environment creating a distinct propagation path. By this characterization, we can use a parametric spatial model \cite{heath_overview_2016} to write the channel through the RIS as 
\begin{equation}\label{rcvdSig_Spatial}
\begin{split}
\mathbf{H}_{\mathrm{RIS}} = \sum_{\ell=1}^{L} \alpha_\ell ~ q_{\ell} ~ \mathbf{a}_R\left( \theta_{R, \ell}, \phi_{R, \ell} \right) ~ \mathbf{a}_{T}^{*}\left( \theta_{T, \ell}, \phi_{T, \ell} \right),
\end{split}
\end{equation}
where %$N_p$ is the number of resolvable paths,
$\alpha_{\ell}$ is a complex scalar representing the $\ell$-th path gain excluding the effects of the RIS element, $q_{\ell}$ is a complex scalar representing the controllable effect of the $\ell$-th RIS element, and $\mathbf{a}_R$ and $\mathbf{a}_T$ represent the array steering vectors at the receiver and the transmitter, respectively, with $\theta$ representing the azimuth angle and $\phi$ representing the elevation angle. 

In general, the amplitude/phase parameter $q_{\ell}$ will be controllable by the RIS, regardless of its implementation technology. Furthermore, a metasurface-based RIS element may also control the angle of the reflection, and hence the angles of arrival at the receiver, $\theta_{R, \ell}$ and $\phi_{R, \ell}$, which will change the receiving array response accordingly. However, changing the reflection angles requires optimizing the entire phase gradient applied by the metasurfaces. In essence, this formulation represents the RIS as a \emph{cluster of reflectors} whose complex gains, and maybe incidence angles, are controllable. Clustered spatial models are known to be highly accurate and are commonly used in wireless standards \cite{almers_survey_2007}. Note that the parameter $q_{\ell}$ is deterministic based on the current configuration of the RIS while the parameters $\alpha_{\ell}$ can be stochastic to model the fading resulting from scattering around the receiver \cite{ayach_spatially_2014, heath_overview_2016}. By this formulation, statistical characterization of small-scale fading at the receiver is possible without forcing the assumption of cascaded fading.

\subsection{Large-scale Path Loss}\label{sec:pathloss}
Another important issue related to channel modeling is how to model the large scale propagation path loss from a transmitter to a receiver through the RIS. Accurate modeling of the path loss through the RIS is critical in assessing the performance of RIS-assisted communication links, especially as compared to other techniques \cite{bjornson_massive_2019}. It is also one possible differentiating aspect between reflectarray and metasurface-based RIS. 

Consider the effect of one RIS element as shown in Fig.~\ref{fig:pathloss}. In the case of a reflectarray-based implementation, each element is a regular antenna whose dimension is typically in the order of $\frac{\lambda}{2}$, while in the case of a metasurface-based RIS, each element is a metasurface tile whose dimensions are orders of magnitudes larger than the wavelength. \emph{Physical size plays an important role in how objects interact with incident electromagnetic waves} \cite{rappaport_wireless_2001}. Smooth objects much larger than the wavelength, e.g., buildings and walls, specularly reflect \emph{the majority of the incident wave}, while objects with dimensions comparable to the wavelength diffusely scatter the incident wave in all directions. This has been observed for a long time in field measurements \cite{rappaport_radio-wave_1994} and is used in site-specific ray-tracing simulations which are known to provide highly accurate results \cite{schaubach_ray_1992}. Furthermore, at millimeter-wave frequencies, the size of objects that can act as reflectors becomes smaller \cite{ben-dor_millimeter-wave_2011}.

In the case of the reflectarray-based RIS, the element acts as a diffuse scatterer. In particular, it only receives \emph{a point on the incident wavefront} and then diffusely scatter it in all directions around the element resulting in further power loss toward the receiver. On the other hand, in the case of a sufficiently large metasurface-based RIS, each element acts as a reflector. It receives \emph{a section of the incident wavefront} and redirects it according to a programmable reflection angle. \emph{There is no further spreading of the wavefront at the metasurface tile.} For this kind of reflection, the reflection angle does not follow Snell's law, i.e., does not equal the incidence angle. It is usually referred to as anomalous reflection \cite{diaz-rubio_generalized_2017}.  

Hence, in the case of the metasurface-based RIS, the path loss \emph{through a single RIS element} will be proportional to the overall distance \cite{schaubach_ray_1992, rappaport_wireless_2001}, $d_{\mathrm{sr}} + d_{\mathrm{rd}}$, i.e.,
\begin{equation}\label{eq:PL_meta}
\begin{split}
\text{PL}_{\mathrm{reflected}} \propto \frac{1}{\left(d_{\mathrm{sr}} + d_{\mathrm{rd}}\right)^n},
\end{split}
\end{equation}
where $n$ is the path loss exponent, e.g. 2 in free space, while in the case of the reflectarray-based RIS, the path loss will be proportional to the product of the distances, $d_{\mathrm{sr}}$ and $d_{\mathrm{rd}}$, i.e.,
\begin{equation}\label{eq:PL_array}
\begin{split}
\text{PL}_{\mathrm{scattered}} \propto \frac{1}{\left(d_{\mathrm{sr}} \times d_{\mathrm{rd}}\right)^n}.
\end{split}
\end{equation}
It is evident that the difference in path loss between the two cases can be immense. 

Before we conclude this section, it is worth mentioning that path loss scaling through the RIS has been a subject of debate in recent literature \cite{ozdogan_pathloss_2019, tang_experi_2019, basar_wireless_2019}. In particular, the assumption that a metasurface acts approximately as a specular reflector is only valid under specific conditions related to physical size of the RIS and \emph{distances to communication endpoints}, i.e., $d_{\mathrm{sr}}$ and $d_{\mathrm{rd}}$. In particular, the RIS physical size has to be large enough relative to the distances, $d_{\mathrm{sr}}$ and $d_{\mathrm{rd}}$, such that the communication endpoints are within the near-field of the RIS. In this case, it is valid to assume that the average received power through the RIS scales with the sum of the distances to the RIS as predicted by~\eqref{eq:PL_meta}. This has been analyzed theoretically in \cite{ozdogan_pathloss_2019, ellingson_path_2019, tang_experi_2019} and observed in empirical measurements conducted in \cite{tang_experi_2019}. In envisioned RIS deployments \cite{renzo_smart_2019}, where an RIS covers large sections of buildings' facades outdoors and walls indoors, it is safe to assume that the transceivers of interest would be in the near-field of the RIS \cite{hu_beyond_2018}, especially in dense networks. For a concrete example, consider an RIS with physical dimensions of  $1.5m \times 1.5m$ deployed on the facade of a building and a transmitted signal at 28 GHz. The near-field of the RIS extends up to $\frac{2 D^2}{\lambda}$ meters away, where $D$ is the largest dimension of the RIS and $\lambda$ the wavelength. Hence, transceivers within $420m$ of the RIS are within its near-field, and path loss through the RIS is well approximated by~\eqref{eq:PL_meta} \cite{tang_experi_2019}.

\section{RIS-assisted Optimization}\label{Sec::Optimization}

The presence of RISs in the propagation environment provides the system designer with vast abilities to alter the wireless channel realization to achieve different objectives in various scenarios. In this section, we provide a brief overview of the state of the art and discuss some of the aspects that differentiate RIS optimization from precoding for traditional MIMO arrays.

\subsection{State-of-the-Art Review}
Recent works have shown the potential of RISs in point-to-point channels \cite{tan_enabling_2018, wu_intelligent_joint, wang_intelligent_2019, basar_transmission_2019, huang_indoor_2019, arun_rfocus:_2019, taha_deep_2019, wu_beamforming_discrete, yang_irs-enhanced_2019, yang_irs_meets_ofdm_2019, you_intelligent_2019, zheng_intelligent_2019, abeywickrama_intelligent_2020, li_towards_LAIA}, downlink broadcast channels \cite{huang_achievable_2018, huang_energy_2018, guo_weighted_2019, huang_reconfigurable_2019, nadeem_asymptotic_2019, nadeem_intelligent_2019, wu_discrete_beamforming_2019, wu_TWC_jointPassiveActive_2019, huang_reconfigurable_2020, di_hybrid_2019}, uplink multiple access channels \cite{cao_delay-constrained_2019, zheng_intelligent_2020, hua_reconfigurable_2019, jiang_over--air_2019}, and even device-to-device interference channels \cite{tan_increasing_2016, welkie_programmable_2017}. In point-to-point, e.g., single-user, scenarios where the receiver has a single antenna, the problem of optimizing the RIS configuration simplifies considerably. In this case, a straightforward strategy is to use the RIS to maximize the effective channel gain, which is equivalent to maximizing the received power or minimizing the transmit power for a given SNR constraint. If the relevant channels are known, then co-phasing all paths is optimal, which can be solved analytically or formulated as a semidefinite program (SDP) \cite{wu_intelligent_joint, wang_intelligent_2019} in case of continuous phase shifts. If phase shifts are discrete, quantizing the continuous solution is near-optimal or a greedy iterative search can be utilized as indicated in \cite{wu_beamforming_discrete}. When the relevant channels are not known, the RIS phase shifts can be optimized based on feedback from the intended receiver, which can be implemented as a beam search procedure \cite{tan_enabling_2018}, or similar algorithms \cite{arun_rfocus:_2019}. Furthermore, when limited channel information or receiver position information can be obtained, they can be used to configure the RIS. In \cite{taha_enabling_2019}, a small subset of the relevant channels is used to infer good RIS configurations based on a deep neural network when the channels are generated using realistic ray-tracing simulation. In \cite{huang_indoor_2019}, a deep neural network was also used but relying on position data rather than partial channel estimates.

In multi-user scenarios, the problem of configuring RIS complicates considerably. In particular, maximizing the channel gains is no longer the only goal since interference must be taken into account. Two heuristic algorithms have been proposed to minimize transmit power under users' SINR constraints in a downlink broadcast scenario in \cite{wu_TWC_jointPassiveActive_2019}, one is based on alternating optimization similar to \cite{wu_intelligent_joint} and the other maximizes the weighted channel gains of all users, where weights are proportional to SINR constraints. In \cite{huang_achievable_2018}, a heuristic algorithm based on majorization-minimization (MM) is used to maximize the sum achievable rate under user rate constraints, when a zero-forcing precoder is applied to cancel all inter-user interference. The same approach is generalized to accommodate energy efficiency maximization in \cite{huang_energy_2018}. Motivated by the fact that the Transmitter-RIS-Receiver link resembles a hybrid beamforming scenario, where the transmitter applies precoding in the digital domain and the RIS applies further analog beamforming using only phase shifts, a hybrid beamforming framework has been proposed in \cite{di_hybrid_2019} to maximize the achievable sum-rate in multi-user downlink setting. Weighted sum-rate maximization has been tackled recently in \cite{guo_weighted_2019}, where efficient suboptimal solutions are found based on fractional programming and alternating direction method of multipliers (ADMM). A deep reinforcement learning (DRL) framework to jointly optimize transmitter precoding and RIS phase shifts has been developed in \cite{huang_reconfigurable_2020} and found to achieve comparable performance to the alternating optimization techniques proposed earlier in the literature.

\subsection{Limited Channel State Information}
In general, the ability to optimize the RIS configuration is limited by available information about the relevant channels. Without at least partial knowledge about the underlying propagation environment, the RIS cannot be used effectively to improve the system performance.
Most recent works on the optimization of RIS parameters have assumed relevant channels are available as side-information, e.g., \cite{wu_intelligent_joint, wu_beamforming_discrete}, and focused on devising algorithms to configure the RIS given channel knowledge. The rationale for this assumption is usually to explore the upper bounds of what can be achieved in practical systems which would have to rely on imperfect estimates of the channels. However, different from traditional wireless systems where channel acquisition is a straightforward matter and could be accounted for by penalizing the achievable rates to account for pilot overhead and SNR degradation from imperfect channel estimation, RIS-assisted wireless systems would encounter more difficulties given the passive nature of the RIS and the massive number of channel parameters to be estimated. Hence, it should be noted that the gains reported for these schemes principally relies on the ability to estimate the channels between the RIS elements and the communication endpoints. Such channel estimation may prove to be tremendously difficult in practice.

The majority of published work on RIS assume a reflectarray-based implementation, where each element is an antenna terminated by a configurable load. As such, RIS elements are neither regular wireless transmitters nor receivers. They do not possess the capability to send pilots or even process pilots to estimate the channel. Adding a receive chain for each element in the RIS is prohibitive. A large number of receive chains, one for each element, would consume a formidable amount of power, and severely limits the attractiveness of RIS assisted communications. Without receive chains, the RIS would not be able to process pilots sent from the two communicating ends to estimate the channel. Moreover, the effect of a single RIS element on the received signal is minuscule; hence, turning the RIS element individually on and off to measure their effects on the received signal is impractical. Next, we present two approaches from the literature to tackle the problem of optimizing the RIS configuration with limited channel information.

The first approach is to forgo channel estimation altogether; instead, the optimization of the RIS can be based on feedback from the receiver. In \cite{tan_enabling_2018}, a three-way beam-searching protocol has been designed to jointly optimize the beam-direction of an IEEE 802.11ad access point and an RIS towards a common receiver. This can be done using a predefined codebook of beam-directions; however, the size of the codebook will be proportional to the number of elements. If there are a large number of RIS elements, training may take a long time \cite{junyi_wang_beam_2009}. This can be a problem in a dynamic environment where the channel coherence time is limited. Nevertheless, relying on feedback from the receiver is a promising solution and the majority of empirically oriented studies have built prototype systems that rely on receiver feedback to guide the optimization of the RIS configuration without explicit knowledge of the channels involved. In \cite{li_towards_LAIA}, the received signal strength indicator (RSSI) is used to judge the current configuration of the RIS and a simple greedy algorithm is devised to sequentially adjust the phase shifts of the reflectarray element-by-element without requiring any additional signaling. In \cite{arun_rfocus:_2019}, RSSI has been also leveraged to guide the optimization of the RIS; however, a simpler two-state phase shifter has been used and a more robust optimization algorithm has been devised to avoid measuring the effect of a single element, which might be very hard in practical scenarios.

The other approach is to equip a \emph{small subset} of the RIS elements with low-power active receivers that are capable of processing pilots and estimating the relevant channels. Hence, the transmitter and the receiver can send pilots to help the RIS acquire \emph{partial information} about the channel. Note that the directly estimated channels constitute a tiny subset of the elements of the RIS and do not provide enough accurate information on their own to facilitate beamforming or other channel manipulations. However, by leveraging tools from compressed sensing and machine learning, the limited available information can be used to compute sufficiently accurate estimates of the channels at all elements, or directly use the partial channel information to infer good RIS configurations. Encouraging results in this direction have been reported in \cite{taha_enabling_2019}, where it has been shown that only 1\% of the RIS elements need to be active in practical propagation scenarios to support a sufficiently accurate prediction of the channels at the remaining elements. This can be attributed to the fact that the channels at all elements result from the interactions of a limited number of scatterers and reflectors in the environment and this limited number of receivers can be regarded as sensors providing descriptors of the environment that carry sufficient information to guide the configuration of the RIS. Note that this technique is somewhat related to \cite{huang_indoor_2019}, where the small number of ``environment descriptors'' can be thought of as providing a spatial fingerprint indicating the position of the user.

The discussion above highlighted the problem of acquiring phase information to optimize the state of the RIS. Indeed, in all currently available prototypes and testbeds \cite{welkie_programmable_2017, tan_increasing_2016, tan_enabling_2018, arun_rfocus:_2019, li_towards_LAIA, dai_reconfigurable_2019}, receiver-side measurements, e.g. signal strength or received power, is used to guide the configuration of the RIS. In these cases, there is a single user and the problem of optimizing the RIS state is more akin to analog beam training. In a sense, the RIS can be thought of as a large beamformer disjoint from the transmitter. One attractive aspect of some of these techniques is that they require no modifications to the used wireless standard. In particular, the RIS training process piggybacks on normal data-carrying transmissions and gradually improves the quality of the received signal without requiring dedicated training signaling. Finally, it is worth noting that these simple proof-of-concept testbeds have used traditional antennas to implement a reflectarray-based RIS and only optimized the phase shifts; it remains to be seen if using metasurfaces to implement the RIS would lead to higher performance gains in real life.

Before we conclude the discussion of channel state information, we would like to point out that the presence of the RIS does not affect the channel estimation procedure between the transmitter and the receiver in the environment. In particular, for any configuration of the RIS, the transmitter and the receiver can estimate the channel between them regularly as if the RIS does not exist. To elaborate, it is straightforward to estimate the overall channel, $\mathbf{H}_{\mathrm{eff}}$, cf.~\eqref{eq:rcvdSig_Cascaded}; however, estimating the channels between the RIS and the two link ends individually or isolating the effect of the RIS is much more challenging \cite{he_cascadedEst_2019}.

\subsection{Optimization Objectives}

 Most recent works on RIS optimization focuses on traditional scenarios similar to those adopted for the study of optimal precoders and receivers for traditional MIMO arrays. However, the presence of the RIS in the middle of the propagation environment provides a unique opportunity to alter the effective channels, cf.~\eqref{eq:rcvdSig_Cascaded}, in ways not possible by traditional MIMO arrays at the communication end-points. Hence, it is possible to pursue optimization objectives that are simply unattainable in traditional systems. For example, as we mentioned earlier in the introduction, millimeter-wave channels are spatially-sparse; in many cases, only one viable propagation path exists which limits the possibility of supporting spatial multiplexing or multi-user MIMO. Most gains in current millimeter-wave systems, e.g. 802.11ad, come from the huge channel bandwidths afforded in these systems. However, given the additive nature of the effects of the RIS, multiple RISs can be deployed across the environment to provide multiple viable propagation paths to increase the rank of the channel and facilitate spatial multiplexing. 
 
Even under the benign assumption of rich scattering underlining the canonical Rayleigh channel model at typical ultra-high frequencies (UHF), i.e., 300 MHz to 3 GHz, ill-conditioned channel realizations, even if full rank, are common, which again limits the spatial multiplexing capability of the channel and decrease the performance of MIMO detectors. The RIS can be used to keep the effective channel well conditioned, which reduces noise enhancement and greatly improves the performance of simple linear MIMO receivers. Preliminary empirical results in this direction were reported in \cite{hougne_optimally_2019}. Another example of these unique RIS optimization objectives was studied in \cite{basar_doppler_2019}, where it has been demonstrated that an RIS can be used to mitigate the effects of Doppler spread and multipath fading. In particular, the effective channel can be made more static over time.
 
Another scenario where an RIS could be immensely useful is in interference channels. In this case, the RIS can be used to facilitate channel sharing by multiple communicating pairs. Assume several single-antenna transmitters are to communicate with corresponding single-antenna receivers and there is an RIS in the middle of the propagation environment. The overall effective channel can be written in the form introduced in Section~\ref{Sec::ChanModel}, where the diagonal terms represent the desired channel and all off-diagonal terms represent interference. The RIS can be configured such that this effective matrix is diagonal allowing spectrum sharing without interference. Empirical results in this direction were reported in \cite{tan_increasing_2016, welkie_programmable_2017}, but so far there are no theoretical results have been reported on the subject.
 
In general, multiple RISs judiciously deployed in the environment give the system operator vast abilities to engineer the propagation environment to satisfy different requirements. Many optimization objectives beyond what is possible using traditional MIMO arrays at the transceivers can be pursued. The presence of the RIS in the middle of the propagation environment can impose many structures on the effective channels depending on the communication scenario. The RIS can operate in multiple modes to satisfy diverse objectives. It can be used as a large analog beamformer to focus power to a single antenna receiver, to enhance the channel rank and condition number to facilitate spatial multiplexing to a multi-antenna receiver or several single-antenna users. It can also be used to diagonalize an interference channel to allow spectrum sharing, or even increase the channel coherence time by mitigating the Doppler spread.

The complexity of optimizing the RIS configuration is evident even for a single RIS aiding a single user transmission especially given the difficulty of acquiring accurate channel information. Multi-user multi-RIS scenarios will certainly be less tractable. This complexity makes machine learning a legitimate avenue for further investigation. Sparse environment sensors can be deployed to monitor the propagation environment and provide information to a centralized controller. This information can be in terms of partial channel coefficients \cite{taha_enabling_2019} or alternatively, user positions \cite{huang_indoor_2019}. Of course, any change in the configuration of the RIS will change the propagation environment and could be sensed and reported back to the controller. This interaction with the environment makes the problem well suited for reinforcement learning techniques. However, the wireless channel is highly dynamic, any movement of the transceivers or even objects in the environment will affect the channel. Hence, any learning techniques will have to operate within the coherence time of the channel and adapt as the channel varies. The vast potential of RIS in shaping the radio environment will be limited only by the ability to find good configurations efficiently within the channel coherence interval.

\section{Numerical Results}\label{Sec::Results}
In this section, we provide numerical results to demonstrate the power of an RIS in shaping a MIMO channel to improve performance while at the same time simplifying precoding at the transmitter and equalization at the receiver. This aspect of the RIS power received little attention in the literature \cite{hougne_optimally_2019}. We consider a simple reflectarray-based RIS and show that by merely adjusting the phase shifts at the RIS, we could effectively orthogonalize the MIMO channel between the transmitter and the receiver and achieve a near-unity condition number. This greatly simplifies the processing at the transmitter and the receiver, shifting the complexity to the RIS controller optimization instead.

We compare the RIS-assisted channel to the canonical Rayleigh fading channel. To focus on the potential gains arising from the improved eigenstructure of the effective channel matrix, we normalize the relevant channels so that the average channel gain is the same regardless of RIS assistance. In particular, we look at the distribution of the condition number, $\kappa \triangleq \frac{|\sigma_{max}|}{|\sigma_{min}|}$ of the channel. The condition number and singular values of a MIMO channel are key properties that characterize the spatial multiplexing capability of the channel. It is well known that for a given channel gain, maximum capacity is achieved when all singular values are equal, which only happens when the channel is orthogonal. Hence, a near-unity condition number, $\kappa$, implies higher capacity. Moreover, an orthogonal channel matrix with equal singular values reduces complexity significantly. For example, uniform power allocation would be optimal, and there is no need to adapt modulation and coding per each spatial stream individually.

\begin{figure}[t] 
\centering
  \subfloat[Rayleigh channel without RIS assistance.]{% 
    \includegraphics[height=7cm,trim={0.5cm 0 0.5cm 0},clip]{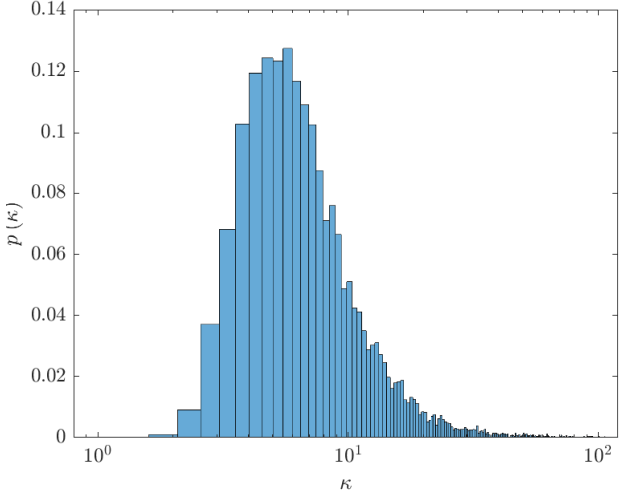} 
  } 
  \\
  \subfloat[RIS-assisted Rayleigh channel.]{% 
  %trim={<left> <lower> <right> <upper>}
    \includegraphics[height=7cm,trim={0.5cm 0 0.5cm 0},clip]{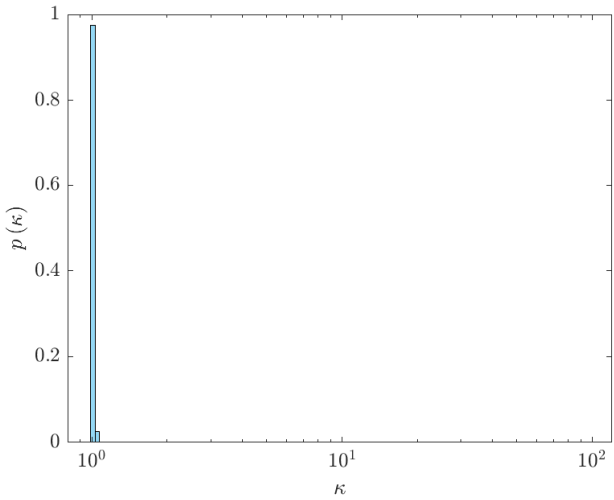} 
  } 
  \centering
  \caption{The effect of the RIS on the distribution of the channel condition number, $\kappa$. Results taken from $10^4$ channel realizations.} 
  \label{fig:condNopt}
\end{figure}

One way to achieve these desirable properties is optimize the RIS configuration to maximize the so called \emph{spectral entropy} of the overall channel, $\mathbf{H}_{\mathrm{eff}}$, c.f.~\eqref{eq:rcvdSig_Cascaded}. The spectral entropy \cite{wenye_yang_coefficient_2005,roy_effective_2007} of an $M \times N$ matrix $\mathbf{X}$, denoted by $\textsf{SE}\left(\mathbf{X}\right)$, is given by
\begin{equation}\label{eq:spectralEntropy}
\begin{split}
\textsf{SE}\left(\mathbf{X}\right) = - \sum_{i} \frac{\sigma_i}{\sum_j \sigma_j} \ln{\left(\frac{\sigma_i}{\sum_j \sigma_j}\right)},
\end{split}
\end{equation}
where $\sigma_i$ is the $i$-th largest singular value of $\mathbf{X}$. Hence, we can write our optimization problem as 
\begin{equation}\label{eq:opt_effrank}
\begin{aligned}
\underset{\boldsymbol{\theta}}{\text{maximize}} \quad & \textsf{SE} \left( \mathbf{H}_{\mathrm{eff}} \right)\\
\text{subject to} \quad & -\pi \le \theta_i \le \pi.
\end{aligned}
\end{equation}
Note that the maximum achievable objective value of~\eqref{eq:opt_effrank} is known to be $\ln{\left(\min{\left(M, N\right)}\right)}$, which only occurs when the channel is orthogonal and can be used to terminate the optimization procedure early. Nevertheless, \eqref{eq:opt_effrank} is not a convex problem. In general, nonconvex problems are NP-Hard and sub-optimal heuristics are necessary to reach good solutions in a reasonable time.

We have applied gradient-based interior-point optimization methods to solve~\eqref{eq:opt_effrank}. Although not convex, the problem seems to be highly amicable to gradient-based techniques. A local-solution with optimal objective, i.e., $SE = \ln{\left(\min{\left(M, N\right)}\right)}$, is typically reached within a few tens of steps regardless of the initial point. 

Consider an RIS-assisted $4 \times 4$ MIMO link where the RIS possess $100$ elements, i.e., $M=N=4$ and $L=100$. In the simulations, QPSK modulation is used over all spatial streams and we have assumed half of the received power came through the RIS and the other half through other paths. This represents a compromise between LOS and NLOS scenarios. We also assume the phase shifts are continuous; however, a limited number of discrete phase shifts are also adequate even if the solution is obtained using continuous optimization then discretized, especially for large RISs. This has been reported in many works, e.g., \cite{wu_beamforming_discrete, wu_discrete_beamforming_2019}, and also proved analytically in \cite{hzhang_reconfigurable_2019}. Fig.~\ref{fig:condNopt} shows the effect of RIS assistance on the channel condition number, $\kappa$. From the figure, a Rayleigh channel without RIS-assistance is very likely to be ill-conditioned; however, with RIS assistance, all channel realizations can be optimized by the RIS to be well-conditioned. This implies that all eigenmodes of the channel are equally good.

Next, we present the effect of optimizing the channel properties on the error performance of the simplest MIMO receiver, the zero-forcing (ZF) linear decoder. The ZF decoder is given by 
\begin{equation}\label{eq:zf_decoder}
\begin{split}
\mathbf{W}_{ZF} = \left( \mathbf{H}_{\mathrm{eff}}^{*} \mathbf{H}_{\mathrm{eff}} \right)^{-1} \mathbf{H}_{\mathrm{eff}}^{*},
\end{split}
\end{equation}
where $\left( \cdot \right)^*$ denotes the Hermitian transpose. Note that in the RIS-assisted case, $\mathbf{H_{\mathrm{eff}}}$ is unitary after optimization and the ZF decoder is equivalent to the optimal ML decoder. Moreover, it is equivalent to the even simpler matched filter receiver given by $\mathbf{H}_{\mathrm{eff}}^{*}$.

\begin{figure}[t] 
  \centering
  \includegraphics[height=7cm,trim={0.5cm 0 0.5cm 0},clip]{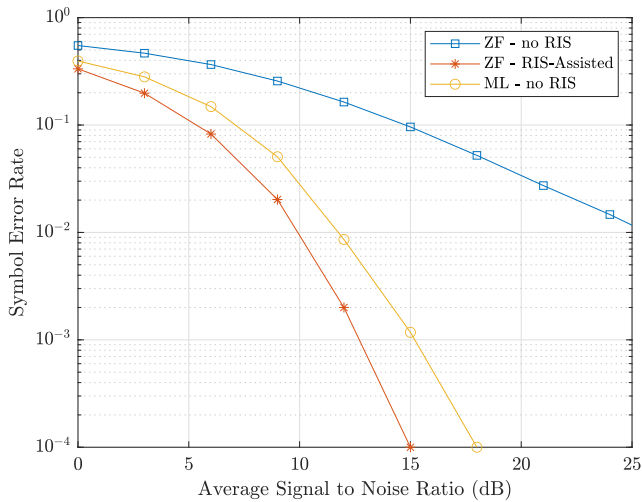} 
  \centering
  \caption{Comparison of the error performance of zero-forcing decoder. The RIS are used to orthogonalize MIMO channels and achieve a unity condition number.} 
  \label{fig:risErrorP}
\end{figure}

Fig.~\ref{fig:risErrorP} shows the average symbol error rate over all four spatial streams. From the figure, RIS assistance has a drastic impact on the error performance of the ZF decoder since noise enhancement is eliminated and ZF decoding is optimal when the channel is orthogonalized. Furthermore, the performance of the ZF decoder in the RIS-assisted channel is actually better than the maximum likelihood decoder in the non-assisted channel. This arises from the fact that in ill-conditioned channels, the weakest spatial stream corresponding to the weakest eigenmode usually suffers from a very low SNR. This problem is overcome in the RIS-assisted scenario since all eigenmodes are equally strong for all channel realizations. Finally, note that we have normalized the channels so that the average channel gain is the same regardless of RIS assistance. In practical scenarios, the RIS can furthermore cause an increase in average channel gain and even higher performance gains are possible.

\section{Future Research Directions}\label{Sec::FutDirec}
In this section, we briefly introduce some potential applications of RIS in wireless networks and discuss potential research directions. 

%Centralized Hybrid Beamforming
\textbf{Centralized Beamforming for IoT Devices:}  The Internet-of-Things (IoT) is an important component in future wireless networks. However, some IoT devices are constrained in both size and energy consumption. Future 5G and beyond cellular networks will operate in the millimeter-wave channels where highly directional antenna gains are essential to achieve reliable high-rate communications. Some IoT devices will be too small in size to support the antenna arrays required to achieve enough beamforming gain to establish a link with a distant base-station. RISs can be used to supply these devices with large beamforming gain, much larger than can be afforded by them given their limited size. Note that both the RIS and the base station will be fixed in place and probably with limited surrounding scatterers, which simplifies the beamforming optimization between the RIS and the base-station. 

% RIS-assisted massive connectivity
\textbf{Experimental Validation of Path Loss Scaling:} Path loss through the RIS will be an important factor in determining the practicality of RIS-assisted communications \cite{bjornson_intelligent_2019}. This is especially important since the RIS is passive and only co-phasing gains are possible. Transparent relaying and backscatter channels are known to suffer from aggravated path loss due to additional spreading of the signal at the relay node or backscatter tag. Traditionally, smooth surfaces much larger than the wavelength are modeled as reflectors\cite{schaubach_ray_1992, rappaport_wireless_2001}, which gives metasurface-based RIS a huge advantage over reflectarray-based RIS. However, this still needs to be validated experimentally. Measurements for $0.35 \lambda$-spaced reflectarray-based RIS were reported in \cite{tang_experi_2019}; however, it is still not clear whether the tighter packing of scatterers in a metasurface-based RIS would improve on these results, especially for the same overall physical size.

\textbf{RF Sensing and Localization:} Another promising direction is RIS-assisted radio-frequency (RF) sensing and localization. The large aperture size of the RIS and its ability to shape the propagation environment can significantly enhance RF sensing capabilities. The channel can be altered to provide favorable conditions for RF sensing then monitored with high accuracy. Encouraging results have been reported in \cite{hu_reconfigurable_2019} with possible applications in energy-efficient surveillance, assisted living, and remote health monitoring. However, the problem of optimizing the configurations of RIS to enhance RF-sensing remains to be studied.

\section{Conclusion}\label{Sec::Conclusion}

Research in RIS-assisted wireless communications is still in its infancy; many practical aspects are not well understood and still need to be thoroughly investigated. However, the potential for this new technology is immense. The idea of being able to change the propagation environment is not only conceptually interesting but also highly beneficial in a variety of scenarios. In this article, we have discussed two candidate implementations for the RIS based on reflectarrays or metasurfaces. We elaborated on channel modeling and how it can be affected by the RIS implementation in terms of channel distribution and large-scale path loss. Furthermore, we presented some of the challenges that need to be addressed in the optimization of RIS-assisted networks. Optimization of the RIS will have to be performed given limited information about the channel and many optimization objectives can be pursued making the problem very complex. In the end, practical optimization techniques are essential to reap the system performance gains resulting from the ability to shape the wireless propagation.

\bibliographystyle{IEEEtran}
\bibliography{main}

\begin{thebibliography}{10}
\providecommand{\url}[1]{#1}
\csname url@samestyle\endcsname
\providecommand{\newblock}{\relax}
\providecommand{\bibinfo}[2]{#2}
\providecommand{\BIBentrySTDinterwordspacing}{\spaceskip=0pt\relax}
\providecommand{\BIBentryALTinterwordstretchfactor}{4}
\providecommand{\BIBentryALTinterwordspacing}{\spaceskip=\fontdimen2\font plus
\BIBentryALTinterwordstretchfactor\fontdimen3\font minus
  \fontdimen4\font\relax}
\providecommand{\BIBforeignlanguage}[2]{{%
\expandafter\ifx\csname l@#1\endcsname\relax
\typeout{** WARNING: IEEEtran.bst: No hyphenation pattern has been}%
\typeout{** loaded for the language `#1'. Using the pattern for}%
\typeout{** the default language instead.}%
\else
\language=\csname l@#1\endcsname
\fi
#2}}
\providecommand{\BIBdecl}{\relax}
\BIBdecl

\bibitem{renzo_smart_2019}
M.~D. Renzo, M.~Debbah, D.-T. Phan-Huy, A.~Zappone, M.-S. Alouini, C.~Yuen,
  V.~Sciancalepore, G.~C. Alexandropoulos, J.~Hoydis, H.~Gacanin, J.~d. Rosny,
  A.~Bounceur, G.~Lerosey, and M.~Fink, ``Smart radio environments empowered by
  reconfigurable {AI} meta-surfaces: an idea whose time has come,''
  \emph{EURASIP J. Wireless Commun. and Netw.}, vol. 2019, no.~1, May 2019.

\bibitem{liang_lisa_2019}
\BIBentryALTinterwordspacing
Y.-C. Liang, R.~Long, Q.~Zhang, J.~Chen, H.~V. Cheng, and H.~Guo, ``Large
  intelligent surface/antennas ({LISA}): Making reflective radios smart,''
  \emph{arXiv:1906.06578 [cs, math]}. [Online]. Available:
  \url{http://arxiv.org/abs/1906.06578}
\BIBentrySTDinterwordspacing

\bibitem{akdeniz_millimeter_2014}
M.~R. Akdeniz, Y.~Liu, M.~K. Samimi, S.~Sun, S.~Rangan, T.~S. Rappaport, and
  E.~Erkip, ``Millimeter {wave} {channel} {modeling} and {cellular} {capacity}
  {evaluation},'' \emph{{IEEE} J. Sel. Areas Commun.}, vol.~32, no.~6, pp.
  1164--1179, Jun. 2014.

\bibitem{ayach_spatially_2014}
O.~E. Ayach, S.~Rajagopal, S.~Abu-Surra, Z.~Pi, and R.~W. Heath, ``Spatially
  sparse precoding in millimeter wave {MIMO} systems,'' \emph{{IEEE} Trans.
  Wireless Commun.}, vol.~13, no.~3, pp. 1499--1513, Mar. 2014.

\bibitem{hum_reconfigurable_2014}
S.~V. Hum and J.~Perruisseau-Carrier, ``Reconfigurable reflectarrays and array
  lenses for dynamic antenna beam control: {A} review,'' \emph{{IEEE} Trans.
  Antennas Propag.}, vol.~62, no.~1, pp. 183--198, Jan. 2014.

\bibitem{he_tunable_2019}
Q.~He, S.~Sun, and L.~Zhou, ``\BIBforeignlanguage{en}{Tunable/{Reconfigurable}
  metasurfaces: {Physics} and applications},''
  \emph{\BIBforeignlanguage{en}{Research}}, vol. 2019, pp. 1--16, Jul. 2019.

\bibitem{arun_rfocus:_2019}
\BIBentryALTinterwordspacing
V.~Arun and H.~Balakrishnan, ``{RFocus}: Practical beamforming for small
  devices,'' \emph{arXiv:1905.05130 [cs]}, May 2019, arXiv: 1905.05130.
  [Online]. Available: \url{http://arxiv.org/abs/1905.05130}
\BIBentrySTDinterwordspacing

\bibitem{chenmetareview16}
H.~Chen, A.~J. Taylor, and N.~Yu, ``A review of metasurfaces: physics and
  applications,'' \emph{Reports on Progress in Physics}, vol.~79, no.~7, Jun.
  2016.

\bibitem{huang_holographic_2019}
\BIBentryALTinterwordspacing
C.~Huang, S.~Hu, G.~C. Alexandropoulos, A.~Zappone, C.~Yuen, R.~Zhang,
  M.~Di~Renzo, and M.~Debbah, ``Holographic {MIMO} surfaces for {6G} wireless
  networks: Opportunities, challenges, and trends,'' \emph{arXiv:1911.12296
  [cs, math]}. [Online]. Available: \url{http://arxiv.org/abs/1911.12296}
\BIBentrySTDinterwordspacing

\bibitem{tan_increasing_2016}
X.~Tan, Z.~Sun, J.~M. Jornet, and D.~Pados, ``Increasing indoor spectrum
  sharing capacity using smart reflect-array,'' in \emph{{IEEE} {Int.} {Conf.}
  on {Commun.} ({ICC})}, Kuala Lumpur, Malaysia, May 2016.

\bibitem{welkie_programmable_2017}
A.~Welkie, L.~Shangguan, J.~Gummeson, W.~Hu, and K.~Jamieson, ``Programmable
  radio environments for smart spaces,'' in \emph{{ACM} {Workshop} on {Hot}
  {Topics} in {Netw.}}, Palo Alto, CA, 2017, pp. 36--42.

\bibitem{hougne_optimally_2019}
P.~d. Hougne, M.~Fink, and G.~Lerosey, ``\BIBforeignlanguage{en}{Optimally
  diverse communication channels in disordered environments with tuned
  randomness},'' \emph{\BIBforeignlanguage{en}{Nature Electronics}}, vol.~2,
  no.~1, pp. 36--41, Jan. 2019.

\bibitem{li_towards_LAIA}
Z.~Li, Y.~Xie, L.~Shangguan, R.~I. Zelaya, J.~Gummeson, W.~Hu, and K.~Jamieson,
  ``Towards programming the radio environment with large arrays of inexpensive
  antennas,'' in \emph{{USENIX} Symp. on Netw. Syst. Des. and Implementation
  ({NSDI})}, Boston, MA, Feb. 2019, pp. 285--300.

\bibitem{heathLozano_foundations_2018}
R.~W. Heath and A.~Lozano, \emph{\BIBforeignlanguage{English}{Foundations of
  {MIMO} Communication}}, 1st~ed.\hskip 1em plus 0.5em minus 0.4em\relax
  Cambridge University Press, Dec. 2018.

\bibitem{griffin_gains_2008}
J.~D. Griffin and G.~D. Durgin, ``Gains {for} {RF} {tags} {using} {multiple}
  {antennas},'' \emph{{IEEE} Trans. Antennas Propag.}, vol.~56, no.~2, pp.
  563--570, Feb. 2008.

\bibitem{elmossallamy_noncoherent_2019}
M.~ElMossallamy, Z.~Han, M.~Pan, R.~Jantti, K.~Seddik, and G.~Y. Li,
  ``Noncoherent {frequency} {shift} {keying} for ambient {backscatter} {over}
  {OFDM} {signals},'' in \emph{{IEEE} {Int.} {Conf.} on {Commun.} ({ICC})},
  Shanghai, China, May 2019.

\bibitem{hzhang_reconfigurable_2019}
H.~Zhang, B.~Di, L.~Song, and Z.~Han, ``Reconfigurable intelligent surfaces
  assisted communications with limited phase shifts: How many phase shifts are
  enough?'' \emph{{IEEE} Trans. Veh. Technol., to be published}.

\bibitem{heath_overview_2016}
R.~W. Heath, N.~González-Prelcic, S.~Rangan, W.~Roh, and A.~M. Sayeed, ``An
  overview of signal processing techniques for millimeter wave mimo systems,''
  \emph{IEEE J. Sel. Topics Signal Process.}, vol.~10, no.~3, pp. 436--453,
  Apr. 2016.

\bibitem{almers_survey_2007}
P.~Almers, E.~Bonek, A.~Burr, N.~Czink, M.~Debbah, V.~Degli-Esposti,
  H.~Hofstetter, P.~Kyösti, D.~Laurenson, G.~Matz, A.~F. Molisch, C.~Oestges,
  and H.~Özcelik, ``Survey of channel and radio propagation models for
  wireless {MIMO} systems,'' \emph{EURASIP J. on Wireless Commun. and Netw.},
  vol. 2007, no.~1, Dec. 2007.

\bibitem{bjornson_massive_2019}
E.~Bj\"{o}rnson, L.~Sanguinetti, H.~Wymeersch, J.~Hoydis, and T.~L. Marzetta,
  ``\BIBforeignlanguage{en}{Massive {MIMO} is a reality{--}{What} is next?:
  {Five} promising research directions for antenna arrays},''
  \emph{\BIBforeignlanguage{en}{Digital Signal Process.}}, vol.~94, pp. 3--20,
  Nov. 2019.

\bibitem{rappaport_wireless_2001}
T.~Rappaport, \emph{Wireless {Communications}: {Principles} and {Practice}},
  2nd~ed.\hskip 1em plus 0.5em minus 0.4em\relax Upper Saddle River, NJ:
  Prentice Hall PTR, 2001.

\bibitem{rappaport_radio-wave_1994}
T.~Rappaport and S.~Sandhu, ``Radio-wave propagation for emerging wireless
  personal-communication systems,'' \emph{{IEEE} Antennas Propag. Mag.},
  vol.~36, no.~5, pp. 14--24, Oct. 1994.

\bibitem{schaubach_ray_1992}
K.~R. Schaubach, N.~J. Davis, and T.~S. Rappaport, ``A ray tracing method for
  predicting path loss and delay spread in microcellular environments,'' in
  \emph{{Veh.} {Technol.} {Soc.} {Conf.} (VTC)}, vol.~2, Denver, CO, May 1992,
  pp. 932--935.

\bibitem{ben-dor_millimeter-wave_2011}
E.~Ben-Dor, T.~S. Rappaport, Y.~Qiao, and S.~J. Lauffenburger,
  ``Millimeter-wave 60 {GHz} outdoor and vehicle {AOA} propagation measurements
  using a broadband channel sounder,'' in \emph{{IEEE} {Global} {Commun.}
  {Conf.} ({GLOBECOM})}, Houston, TX, Dec. 2011.

\bibitem{diaz-rubio_generalized_2017}
A.~D\'iaz-Rubio, V.~S. Asadchy, A.~Elsakka, and S.~A. Tretyakov, ``From the
  generalized reflection law to the realization of perfect anomalous
  reflectors,'' \emph{Science Advances}, vol.~3, no.~8, Aug. 2017.

\bibitem{ozdogan_pathloss_2019}
\BIBentryALTinterwordspacing
O.~\"{O}zdogan, E.~Bj\"{o}rnson, and E.~G. Larsson, ``Intelligent reflecting
  surfaces: Physics, propagation, and pathloss modeling,'' \emph{IEEE Wireless
  Commun. Lett., to be published}. [Online]. Available:
  \url{http://arxiv.org/abs/1911.03359}
\BIBentrySTDinterwordspacing

\bibitem{tang_experi_2019}
\BIBentryALTinterwordspacing
W.~Tang, M.~Z. Chen, X.~Chen, J.~Y. Dai, Y.~Han, M.~Di~Renzo, Y.~Zeng, S.~Jin,
  Q.~Cheng, and T.~J. Cui, ``Wireless communications with reconfigurable
  intelligent surface: {Path} loss modeling and experimental measurement,''
  \emph{arXiv:1911.05326 [cs, eess, math]}, Nov. 2019. [Online]. Available:
  \url{http://arxiv.org/abs/1911.05326}
\BIBentrySTDinterwordspacing

\bibitem{basar_wireless_2019}
\BIBentryALTinterwordspacing
E.~Basar, M.~Di~Renzo, J.~de~Rosny, M.~Debbah, M.-S. Alouini, and R.~Zhang,
  ``Wireless {Communications} {Through} {Reconfigurable} {Intelligent}
  {Surfaces},'' \emph{arXiv:1906.09490 [cs, eess, math]}, Jun. 2019, arXiv:
  1906.09490. [Online]. Available: \url{http://arxiv.org/abs/1906.09490}
\BIBentrySTDinterwordspacing

\bibitem{ellingson_path_2019}
\BIBentryALTinterwordspacing
S.~W. Ellingson, ``Path loss in reconfigurable intelligent surface-enabled
  channels,'' \emph{arXiv:1912.06759 [cs, eess, math]}. [Online]. Available:
  \url{http://arxiv.org/abs/1912.06759}
\BIBentrySTDinterwordspacing

\bibitem{hu_beyond_2018}
S.~Hu, F.~Rusek, and O.~Edfors, ``\BIBforeignlanguage{en}{Beyond massive
  {MIMO}: The potential of data transmission with large intelligent
  surfaces},'' \emph{\BIBforeignlanguage{en}{{IEEE} Trans. Signal Process.}},
  vol.~66, no.~10, pp. 2746--2758, May 2018.

\bibitem{tan_enabling_2018}
X.~Tan, Z.~Sun, D.~Koutsonikolas, and J.~M. Jornet, ``Enabling indoor mobile
  millimeter-wave networks based on smart reflect-arrays,'' in \emph{{IEEE}
  {Conf.} on {Comput.} {Commun.} {(INFOCOM)}}, Honolulu, HI, Apr. 2018, pp.
  270--278.

\bibitem{wu_intelligent_joint}
Q.~Wu and R.~Zhang, ``Intelligent {reflecting} {surface} {enhanced} {wireless}
  {network}: {Joint} {active} and {passive} {beamforming} {design},'' in
  \emph{{IEEE} {Global} {Commun.} {Conf.} ({GLOBECOM})}, Abu Dhabi, UAE, Dec.
  2018.

\bibitem{wang_intelligent_2019}
\BIBentryALTinterwordspacing
P.~Wang, J.~Fang, X.~Yuan, Z.~Chen, H.~Duan, and H.~Li, ``Intelligent
  reflecting surface-assisted millimeter wave communications: Joint active and
  passive precoding design,'' \emph{arXiv:1908.10734 [eess]}. [Online].
  Available: \url{http://arxiv.org/abs/1908.10734}
\BIBentrySTDinterwordspacing

\bibitem{basar_transmission_2019}
E.~Basar, ``Transmission through large intelligent surfaces: {A} new frontier
  in wireless communications,'' in \emph{{European} {Conf.} on {Netw.} and
  {Commun.} ({EuCNC})}, Valencia, Spain, Jun. 2019, pp. 112--117.

\bibitem{huang_indoor_2019}
\BIBentryALTinterwordspacing
C.~Huang, G.~C. Alexandropoulos, C.~Yuen, and M.~Debbah, ``Indoor signal
  focusing with deep learning designed reconfigurable intelligent surfaces,''
  in \emph{IEEE Int. Workshop on Signal Process. Adv. in Wireless Commun.
  (SPAWC)}, Cannes, France, Jul. 2019. [Online]. Available:
  \url{http://arxiv.org/abs/1905.07726}
\BIBentrySTDinterwordspacing

\bibitem{taha_deep_2019}
A.~Taha, M.~Alrabeiah, and A.~Alkhateeb, ``Deep learning for large intelligent
  surfaces in millimeter wave and massive {MIMO} systems,'' in \emph{{IEEE}
  {Global} {Commun.} {Conf.} ({GLOBECOM})}, Waikoloa, HI, Dec. 2019.

\bibitem{wu_beamforming_discrete}
Q.~Wu and R.~Zhang, ``Beamforming {optimization} for {intelligent} {reflecting}
  {surface} with {discrete} {phase} {shifts},'' in \emph{{IEEE} {Int.} {Conf.}
  on {Acoust.}, {Speech} and {Signal} {Process.} ({ICASSP})}, Brighton, United
  Kingdom, May 2019, pp. 7830--7833.

\bibitem{yang_irs-enhanced_2019}
Y.~Yang, S.~Zhang, and R.~Zhang, ``{IRS}-enhanced {OFDM}: Power allocation and
  passive array optimization,'' Dec. 2019.

\bibitem{yang_irs_meets_ofdm_2019}
\BIBentryALTinterwordspacing
Y.~Yang, B.~Zheng, S.~Zhang, and R.~Zhang, ``Intelligent reflecting surface
  meets {OFDM}: Protocol design and rate maximization,'' \emph{arXiv:1906.09956
  [cs, math]}. [Online]. Available: \url{http://arxiv.org/abs/1906.09956}
\BIBentrySTDinterwordspacing

\bibitem{you_intelligent_2019}
\BIBentryALTinterwordspacing
C.~You, B.~Zheng, and R.~Zhang, ``Intelligent reflecting surface with discrete
  phase shifts: Channel estimation and passive beamforming,''
  \emph{arXiv:1911.03916 [cs, math]}, Nov. 2019. [Online]. Available:
  \url{http://arxiv.org/abs/1911.03916}
\BIBentrySTDinterwordspacing

\bibitem{zheng_intelligent_2019}
B.~Zheng and R.~Zhang, ``Intelligent reflecting surface-enhanced ofdm: Channel
  estimation and reflection optimization,'' \emph{{IEEE} Wireless Commun.
  Lett.}, vol.~9, no.~4, Apr. 2020.

\bibitem{abeywickrama_intelligent_2020}
\BIBentryALTinterwordspacing
S.~Abeywickrama, R.~Zhang, and C.~Yuen, ``Intelligent reflecting surface:
  Practical phase shift model and beamforming optimization,''
  \emph{arXiv:1907.06002 [cs, eess, math]}. [Online]. Available:
  \url{http://arxiv.org/abs/1907.06002}
\BIBentrySTDinterwordspacing

\bibitem{huang_achievable_2018}
C.~Huang, A.~Zappone, M.~Debbah, and C.~Yuen, ``Achievable rate maximization by
  passive intelligent mirrors,'' in \emph{{IEEE} {Int.} {Conf.} on {Acoust.},
  {Speech} and {Signal} {Process.} ({ICASSP})}, Apr. 2018, pp. 3714--3718.

\bibitem{huang_energy_2018}
C.~Huang, G.~C. Alexandropoulos, A.~Zappone, M.~Debbah, and C.~Yuen, ``Energy
  efficient multi-user {MISO} communication using low resolution large
  intelligent surfaces,'' in \emph{{IEEE} {Globecom} {Workshops} ({GC}
  {Wkshps})}, Abu Dhabi, UAE, Dec. 2018.

\bibitem{guo_weighted_2019}
H.~Guo, Y.-C. Liang, J.~Chen, and E.~G. Larsson, ``Weighted sum-rate
  maximization for intelligent reflecting surface enhanced wireless networks,''
  in \emph{{IEEE} {Global} {Commun.} {Conf.} ({GLOBECOM})}, Waikoloa, HI, Dec.
  2019.

\bibitem{huang_reconfigurable_2019}
C.~Huang, A.~Zappone, G.~C. Alexandropoulos, M.~Debbah, and C.~Yuen,
  ``Reconfigurable intelligent surfaces for energy efficiency in wireless
  communication,'' \emph{{IEEE} Trans. Wireless Commun.}, vol.~18, no.~8, pp.
  4157--4170, Aug. 2019.

\bibitem{nadeem_asymptotic_2019}
\BIBentryALTinterwordspacing
Q.-U.-A. Nadeem, A.~Kammoun, A.~Chaaban, M.~Debbah, and M.-S. Alouini,
  ``Asymptotic max-min {SINR} analysis of reconfigurable intelligent surface
  assisted {MISO} systems,'' \emph{arXiv:1903.08127 [cs, math]}. [Online].
  Available: \url{http://arxiv.org/abs/1903.08127}
\BIBentrySTDinterwordspacing

\bibitem{nadeem_intelligent_2019}
\BIBentryALTinterwordspacing
------, ``Intelligent reflecting surface assisted wireless communication:
  Modeling and channel estimation,'' \emph{arXiv:1906.02360 [cs, eess, math]}.
  [Online]. Available: \url{http://arxiv.org/abs/1906.02360}
\BIBentrySTDinterwordspacing

\bibitem{wu_discrete_beamforming_2019}
Q.~Wu and R.~Zhang, ``Beamforming optimization for wireless network aided by
  intelligent reflecting surface with discrete phase shifts,'' \emph{IEEE
  Trans. on Commun., to be published}.

\bibitem{wu_TWC_jointPassiveActive_2019}
------, ``Intelligent reflecting surface enhanced wireless network via joint
  active and passive beamforming,'' \emph{{IEEE} Trans. Wireless Commun.},
  vol.~18, no.~11, pp. 5394--5409, Nov. 2019.

\bibitem{huang_reconfigurable_2020}
\BIBentryALTinterwordspacing
C.~Huang, R.~Mo, and C.~Yuen, ``Reconfigurable intelligent surface assisted
  multiuser {MISO} systems exploiting deep reinforcement learning,''
  \emph{{IEEE} J. Sel. Areas Commun., to be published}. [Online]. Available:
  \url{http://arxiv.org/abs/2002.10072}
\BIBentrySTDinterwordspacing

\bibitem{di_hybrid_2019}
\BIBentryALTinterwordspacing
B.~Di, H.~Zhang, L.~Song, Y.~Li, Z.~Han, and H.~V. Poor, ``Hybrid {Beamforming}
  for {Reconfigurable} {Intelligent} {Surface} based {Multi}-user
  {Communications}: {Achievable} {Rates} with {Limited} {Discrete} {Phase}
  {Shifts},'' \emph{arXiv:1910.14328 [cs, eess, math]}, Oct. 2019, arXiv:
  1910.14328. [Online]. Available: \url{http://arxiv.org/abs/1910.14328}
\BIBentrySTDinterwordspacing

\bibitem{cao_delay-constrained_2019}
\BIBentryALTinterwordspacing
Y.~Cao and T.~Lv, ``Delay-constrained joint power control, user detection and
  passive beamforming in intelligent reflecting surface assisted uplink mmwave
  system,'' \emph{arXiv:1912.10030 [eess]}. [Online]. Available:
  \url{http://arxiv.org/abs/1912.10030}
\BIBentrySTDinterwordspacing

\bibitem{zheng_intelligent_2020}
\BIBentryALTinterwordspacing
B.~Zheng, C.~You, and R.~Zhang, ``Intelligent reflecting surface assisted
  multi-user {OFDMA}: Channel estimation and training design,''
  \emph{arXiv:2003.00648 [cs, math]}. [Online]. Available:
  \url{http://arxiv.org/abs/2003.00648}
\BIBentrySTDinterwordspacing

\bibitem{hua_reconfigurable_2019}
\BIBentryALTinterwordspacing
S.~Hua, Y.~Zhou, K.~Yang, and Y.~Shi, ``Reconfigurable intelligent surface for
  green edge inference,'' \emph{arXiv:1912.00820 [cs, eess, math]}. [Online].
  Available: \url{http://arxiv.org/abs/1912.00820}
\BIBentrySTDinterwordspacing

\bibitem{jiang_over--air_2019}
\BIBentryALTinterwordspacing
T.~Jiang and Y.~Shi, ``Over-the-air computation via intelligent reflecting
  surfaces,'' \emph{arXiv:1904.12475 [cs, eess, math]}. [Online]. Available:
  \url{http://arxiv.org/abs/1904.12475}
\BIBentrySTDinterwordspacing

\bibitem{taha_enabling_2019}
\BIBentryALTinterwordspacing
A.~Taha, M.~Alrabeiah, and A.~Alkhateeb, ``Enabling large intelligent surfaces
  with compressive sensing and deep learning,'' \emph{arXiv:1904.10136 [cs,
  eess, math]}, Apr. 2019, arXiv: 1904.10136. [Online]. Available:
  \url{http://arxiv.org/abs/1904.10136}
\BIBentrySTDinterwordspacing

\bibitem{junyi_wang_beam_2009}
J.~Wang, Z.~Lan, C.~Pyo, T.~Baykas, C.~Sum, M.~Rahman, J.~Gao, R.~Funada,
  F.~Kojima, H.~Harada, and S.~Kato, ``Beam codebook based beamforming protocol
  for multi-{Gbps} millimeter-wave {WPAN} systems,'' \emph{{IEEE} J. Sel. Areas
  Commun.}, vol.~27, no.~8, pp. 1390--1399, Oct. 2009.

\bibitem{dai_reconfigurable_2019}
\BIBentryALTinterwordspacing
L.~Dai, B.~Wang, M.~Wang, X.~Yang, J.~Tan, S.~Bi, S.~Xu, F.~Yang, Z.~Chen,
  M.~Di~Renzo, and L.~Hanzo, ``Reconfigurable intelligent surface-based
  wireless communication: Antenna design, prototyping and experimental
  results,'' \emph{arXiv:1912.03620 [cs, math]}. [Online]. Available:
  \url{http://arxiv.org/abs/1912.03620}
\BIBentrySTDinterwordspacing

\bibitem{he_cascadedEst_2019}
Z.-Q. He and X.~Yuan, ``Cascaded channel estimation for large intelligent
  metasurface assisted massive mimo,'' \emph{IEEE Wireless Commun. Lett., to be
  published}.

\bibitem{basar_doppler_2019}
\BIBentryALTinterwordspacing
E.~Basar and I.~F. Akyildiz, ``Reconfigurable intelligent surfaces for
  {Doppler} effect and multipath fading mitigation,'' \emph{arXiv:1912.04080
  [cs, eess, math]}. [Online]. Available: \url{http://arxiv.org/abs/1912.04080}
\BIBentrySTDinterwordspacing

\bibitem{wenye_yang_coefficient_2005}
W.~Yang, J.~Gibson, and T.~He, ``Coefficient rate and lossy source coding,''
  \emph{{IEEE} Trans. Inf. Theory}, vol.~51, no.~1, pp. 381--386, Jan. 2005.

\bibitem{roy_effective_2007}
O.~Roy and M.~Vetterli, ``The effective rank: {A} measure of effective
  dimensionality,'' in \emph{{European} {Signal} {Process.} {Conf.}}, Poznan,
  Poland, Sep. 2007, pp. 606--610.

\bibitem{bjornson_intelligent_2019}
\BIBentryALTinterwordspacing
E.~Bj\"{o}rnson, O.~\"{O}zdogan, and E.~G. Larsson, ``Intelligent reflecting
  surface vs. decode-and-forward: How large surfaces are needed to beat
  relaying?'' \emph{arXiv:1906.03949 [cs, math]}, Jun. 2019. [Online].
  Available: \url{http://arxiv.org/abs/1906.03949}
\BIBentrySTDinterwordspacing

\bibitem{hu_reconfigurable_2019}
\BIBentryALTinterwordspacing
J.~Hu, H.~Zhang, B.~Di, L.~Li, L.~Song, Y.~Li, Z.~Han, and H.~V. Poor,
  ``Reconfigurable intelligent surfaces based {RF} sensing: {Design},
  optimization, and implementation,'' \emph{arXiv:1912.09198 [cs, eess]}, Dec.
  2019. [Online]. Available: \url{http://arxiv.org/abs/1912.09198}
\BIBentrySTDinterwordspacing

\end{thebibliography}

\end{document}